\documentclass[aps,12pt,nofootinbib,showpacs]{revtex4-1}

\usepackage[english]{babel}
\usepackage{amsmath}
\usepackage{amssymb}
\usepackage{amsbsy}
\usepackage{amstext}
\usepackage{graphicx}
\usepackage{subfigure}
\usepackage{bm}
\usepackage{color}
\usepackage{verbatim}

\newcommand{\be}{\begin{eqnarray}}
\newcommand{\ee}{\end{eqnarray}}
\newcommand{\bdm}{\begin{displaymath}}
\newcommand{\edm}{\end{displaymath}}
\newcommand{\ds}{\displaystyle}
\newcommand{\ba}{\begin{array}}
\newcommand{\ea}{\end{array}}
\newcommand{\pa}[1]{\left(#1\right)}
\newcommand{\paq}[1]{\left[#1\right]}

\newcommand{\Pp}{{\bf p}}

\newcommand{\X}{{\bf x}}

\begin{document}

\title{Conservative dynamics of binary systems to fourth Post-Newtonian order in the EFT approach I: Regularized Lagrangian}

\author{Stefano Foffa$^{\rm 1}$ and Riccardo Sturani$^{\rm 2}$}

\affiliation{$(1)$ D\'epartement de Physique Th\'eorique and Centre for Astroparticle Physics, Universit\'e de 
             Gen\`eve, CH-1211 Geneva, Switzerland\\
             $(2)$ International Institute of Physics, Universidade Federal do Rio Grande do Norte, Campus Universit\'ario, Lagoa Nova, Natal-RN 59078-970, Brazil}

\email{stefano.foffa@unige.ch, riccardo@iip.ufrn.br}

\begin{abstract}
Within the framework of the effective field theory approach to the post-Newtonian approximation to General Relativity,
we report the so far missing $G^3$ and $G^4$  sectors of the conservative two body
dynamics at fourth perturbative order for non-spinning compact objects.
Following the standard procedure, the exchange of gravitational interaction is
integrated out to obtain a systematic expansion in terms of
Feynman graphs.
Building on this result and on previously obtained ones, we are able 
to write the local-in-time regularized Lagrangian at fourth
post-Newtonian order, which is in agreement with results previously computed
with other methods.
\end{abstract}



\maketitle

\section{Introduction}
\label{sec:intro}
The present paper reports the $G^3$ and the local-in-time $G^4$ sectors of the
near zone Lagrangian governing the
relativistic conservative dynamics of the two non-spinning body motion at fourth
post-Newtonian (PN) approximation order to General Relativity (GR), concluding the effort initiated by the
authors and collaborators with the derivation of the third PN order \cite{Foffa:2011ub}
and then with the partial fourth PN results for the $G, G^2$ \cite{Foffa:2012rn}
and $G^5$ sectors \cite{Foffa:2016rgu} by using Effective Field Theory methods
(EFT) applied to gravity \cite{Goldberger:2004jt},
see \cite{Goldberger:2007hy,Foffa:2013qca,Porto:2016pyg,Levi:2018nxp} for reviews,
being $G$ Newton's constant (in 3+1 dimensions).
At generic nPN order the dynamics is naturally split in sectors of the type
$G^j v^{2(n-j+1)}$, with $0<j\leq n+1$,
as $v^2\sim GM/r$ by using Kepler law, with $M$ being the total mass of the
binary system. The velocity $v$ is the expansion parameter, two successive PN
orders being separated by a factor $v^2$.
We work in dimensional regularization and keep our expressions in $d$ dimensions,
without distinguishing the nature of poles as $d\to 3$.
Following procedure outlined in \cite{Manohar:2006nz,Porto:2017dgs},
the companion paper \cite{Foffa:2019yfl} will complete the project by properly identifying the
intermediate infrared (IR) and ultraviolet (UV) divergences, as well as providing
a self-contained, ambiguity-free renormalization of the effective theory, including also conservative contributions from radiation modes.

The two body problem in GR has been studied at length in the last decades, see \cite{Blanchet:2013haa,Futamase:2007zz,Schafer:2018kuf,Goldberger:2007hy,Foffa:2013qca,Porto:2016pyg},
and its phenomenological relevance relies on being an unavoidable ingredient
for constructing template waveforms \cite{Taracchini:2012ig,Schmidt:2014iyl}
necessary for the detections of gravitational waves made by the
laser interferometric detector LIGO \cite{TheLIGOScientific:2016pea,Abbott:2017vtc,Abbott:2017gyy,LIGOScientific:2018jsj}
and jointly with his European homologous Virgo \cite{Abbott:2017oio,TheLIGOScientific:2017qsa}.
Solving this problem has further applications to numerical relativity
\cite{Mroue:2013xna},
self force calculations \cite{Akcay:2015pza}, which are also related to  waveform template construction, and in general the GR two-body dynamics is a problem
rich of intriguing theoretical aspects, representing an highly non-trivial test-bed for classical field theory beyond its phenomenological applications.

Our work is developed within the framework of EFT methods,
while other groups have obtained the complete 4PN conservative
dynamics with mutually consistent results using different methods:
within the Arnowitt-Deser-Misner (ADM) Hamiltonian formalism \cite{Damour:2014jta,Damour:2015isa,Jaranowski:2015lha,Damour:2016abl}
and within the Fokker action one in 
\cite{Bernard:2015njp,Bernard:2016wrg,Bernard:2017bvn,Marchand:2017pir,Bernard:2017ktp}.

To make contact with standard literature, we will refer to
{\it near-zone} dynamics as the one involving the interaction of massive bodies at the scale of their mutual distance $r$, mediated by
{\it off-shell}, "potential" gravitational modes, with 4-momentum $k^\mu$ scaling as $(k_0,k)\sim(1/r,v/r)$, being $v$ the relative velocity.
On the other hand, we will refer to {\it far zone} as the radiation region, which
is at a distance from the massive bodies equal or larger than the radiation wavelength
$\sim r/v$, and where the dynamics involves {\it on-shell}, radiative gravitational modes with $(k_0,k)\sim(v/r,v/r)$.

The outline of the paper is as follows.
In sec.~\ref{sec:method} we give an overview of the methods employed in deriving the
GR two-body effective Lagrangian in the PN approximation, the key ingredient being the perturbative functional
integration of potential gravitational modes in the near zone; in the same section we also recall
the invariance of gauge independent physical results under
manipulation of (gauge dependent) Lagrangian terms by use of the equations of
motion.

In sec.~\ref{sec:results} we show the bare result of the Feynman diagram computation at order $G^3$ and $G^4$, while revisiting also the 3PN sector.
In sec.~\ref{sec:reg_lag} we combine these
results with those of \cite{Foffa:2012rn, Foffa:2016rgu}  and we re-organize them for ease of comparison with previous works.

While in this paper we are concerned with the
contribution from potential modes to the near zone
Lagrangian, for directly compare with previous
literature
we will simply add to our computation the far-zone conservative and local-in-time
contribution due to the tail, postponing to \cite{Foffa:2019yfl} a
discussion for the theoretical foundations for this
procedure, which relies on the proper identification
of IR/UV poles.
In particular we show how this procedure enables the decomposition of the final
result into two structures: a first one, containing all the remaining poles, which is irrelevant for the dynamics because it vanishes on the
equations of motion; and a second, finite one, which we show being equivalent to the Lagrangian reported
in \cite{Bernard:2017ktp}.
Finally, sec.~\ref{sec:conclusions} gathers our conclusions.

\section{Method}
\label{sec:method}
We describe point particles via a {\it world-line}
action $\mathcal{S}_{pp}$ which resides on the
world-line of the binary constituents and a {\it bulk} part $\mathcal{S}_{EH+GF}$
which encodes the dynamics of gravity:
\be
\label{eq:spp}
\mathcal{S}_{pp}=-\sum_{a=1,2}m_a\int d\tau_a=-\sum_{a=1,2}m_a\int
\sqrt{-g_{\mu\nu}(x_a)dx^\mu_a dx^\nu_a}\,,
\ee
\be
\label{eq:sEH}
{\mathcal S}_{EH+GF}=2\Lambda^2\int d^{d+1}x\sqrt{-g}\pa{R(g)-\frac 12\Gamma^\mu\Gamma_\mu}\,,
\ee
where the gauge fixing term with $\Gamma^\mu\equiv g^{\nu\rho}\Gamma^\mu_{\nu\rho}$
has been added to the standard Ricci term of the Einstein-Hilbert action.
Here $\Lambda^{-2}\equiv 32\pi G^{(d)}$ is defined in terms of the $d-$dimensional
Newton's constant $G^{(d)}=G(\sqrt{4\pi}e^{\gamma_E/2}L_0)^{d-3}$ being $L_0$ an arbitrary length scale introduced to adjust dimensions,
with the presence of $\pi$ and Euler Gamma factors introduced in
its definition for later convenience.
The introduction of the arbitrary scale $L_0$ is made necessary by the use of dimensional
regularization, but this length-scale does not appear in expressions involving physical observables. 

As suggested in \cite{Kol:2007bc} and already done in our previous papers reporting 3PN and partial
4PN results, we find convenient to adopt the metric parametrization
\be
\label{eq:ansatz}
g_{\mu\nu}=e^{2\phi/\Lambda}\pa{
\ba{cc}
\ds -1 &\ds \frac{A_j}\Lambda\\\ds
\frac{A_i}\Lambda &\ds e^{-c_d\phi/\Lambda}\pa{\delta_{ij}+\frac{\sigma_{ij}}\Lambda}-
\frac{A_iA_j}{\Lambda^2}
\ea}
\ee
with $c_d\equiv 2\frac{d-1}{d-2}$ and Latin indices $i,j$ runs over spatial indices
only. This ansatz has the virtue of diagonalizing the kinetic terms of the fields
$\phi,A_i,\sigma_{ij}$ speeding up the computations.

Substituting ansatz (\ref{eq:ansatz}) into actions (\ref{eq:spp},\ref{eq:sEH})
one can rewrite $\mathcal{S}_{pp}$ as
\renewcommand{\arraystretch}{1.4}
\be
\label{eq:spp_KK}
{\mathcal S}_{pp}=-\sum_{a=1,2}m_a\ds \int {\rm d}\tau_a = \ds-\sum_{a=1,2}m_a
\int {\rm d}t_a\ e^{\phi/\Lambda}
\sqrt{\pa{1-\frac{A_i}{\Lambda}v^i_a}^2
-e^{-c_d \phi/\Lambda}\pa{v^2_a+\frac{\sigma_{ij}}{\Lambda} v^i_av^j_a}}\,,
\ee
\renewcommand{\arraystretch}{1.4}
and its Taylor expansion provides the various particle-gravity vertices of the EFT.

Analogously we report ${\mathcal S}_{bulk}$ truncated at the order relevant for the computation
in this paper
\renewcommand{\arraystretch}{1.4}
\be
\label{eq:sEH_KK}
\ba{rcl}
\ds {\mathcal S}_{bulk} &\supset &\ds \int {\rm d}^{d+1}x\sqrt{-\gamma}
\left\{\frac{1}{4}\left[(\vec{\nabla}\sigma)^2-2(\vec{\nabla}\sigma_{ij})^2-\left(\dot{\sigma}^2-2(\dot{\sigma}_{ij})^2\right){\rm e}^{\frac{-c_d \phi}{\Lambda}}\right]- c_d \left[(\vec{\nabla}\phi)^2-\dot{\phi}^2 {\rm e}^{-\frac{c_d\phi}{\Lambda}}\right]\right.\\
&&\ds
+\left[\frac{F_{ij}^2}{2}+\left(\vec{\nabla}\!\!\cdot\!\!\vec{A}\right)^2 -\dot{\vec{A}}^2 {\rm e}^{-\frac{c_d\phi}{\Lambda}} \right]
{\rm e}^{\frac{c_d \phi}{\Lambda}}+\frac 2\Lambda\paq{\pa{F_{ij}A^i\dot{A^j}+\vec{A}\!\!\cdot\!\!\dot{\vec{A}}(\vec{\nabla}\!\!\cdot\!\!\vec{A})}
{\rm e}^{\frac{c_d \phi}{\Lambda}}-c_d\dot{\phi}\vec{A}\!\!\cdot\!\!\vec{\nabla}\phi}
\\
&&\ds
+2 c_d \left(\dot{\phi}\vec{\nabla}\!\!\cdot\!\!\vec{A}-\dot{\vec{A}}\!\!\cdot\!\!\vec{\nabla}\phi\right)
+\frac{\dot{\sigma}_{ij}}{\Lambda}\left(-\delta^{ij}A_l\hat{\Gamma}^l_{kk}+ 2A_k\hat{\Gamma}^k_{ij}-2A^i\hat{\Gamma}^j_{kk}\right)-c_d\frac{\dot{\phi}^2\vec{A}^2}{\Lambda^2}\\
&&\ds
-\left.\frac{1}{\Lambda}\left(\frac{\sigma}{2}\delta^{ij}-\sigma^{ij}\right)
\left({\sigma_{ik}}^{,l}{\sigma_{jl}}^{,k}-{\sigma_{ik}}^{,k}{\sigma_{jl}}^{,l}+\sigma_{,i}{\sigma_{jk}}^{,k}-\sigma_{ik,j}\sigma^{,k}
\right)\right\}\,,
\ea
\ee
\renewcommand{\arraystretch}{1.}
where $F_{ij}\equiv A_{j,i}- A_{i,j}$, and $\hat{\Gamma}^i_{jk}$ is the connection of the purely spatial $d$-dimensional metric $\gamma_{ij}\equiv\delta_{ij}+\sigma_{ij}/\Lambda$,
which is also used above to raise and contract spatial indices.
All spatial derivatives are understood as simple (not covariant) derivatives
and when ambiguities might raise gradients are always meant to act on
contravariant fields, e.g.
$\vec{\nabla}\!\!\cdot\!\!\vec{A}\equiv\gamma^{ij}A_{i,j}$
and $F_{ij}^2\equiv\gamma^{ik}\gamma^{jl}F_{ij}F_{kl}$.

The 2-body effective action is found by integrating out the gravity fields from the
${\mathcal S}_{bulk}+{\mathcal S}_{pp}$
\begin{equation}
\label{eq:seff}
 \exp[\text{i}\mathcal{S}_{eff}]=\int D\phi DA_i D\sigma_{ij} \exp[\text{i}
({\mathcal S}_{bulk}+ {\mathcal S}_{pp})]\,.
\end{equation}
As usual in field theory, the functional integration can be perturbatively 
expanded in terms of Feynman diagrams involving the gravitational 
degrees of freedom as internal lines only. No gravitational modes will appear among
external lines, with the massive bodies playing the role of non-dynamical sources and sinks
of gravitational modes; only diagrams corresponding to classical
contributions to the effective Lagrangian need to be considered.

To make manifest the $v$ scaling necessary to classify the results according to the PN
expansion, it is convenient to work with the a mixed decomposition of the fields
working in direct time-coordinate and Fourier transforming on the space ones:
\renewcommand{\arraystretch}{1.4}
\be
\label{Fourk}
W^a_p(t)  \equiv \ds\int {\rm d}^dx\, W^a(t,x) e^{-\text{i} p\cdot x}\,\quad
 {\rm with\ } W^a=\{\phi,A_i,\sigma_{ij}\}\,.
\ee
\renewcommand{\arraystretch}{1}
The amplitudes corresponding to each diagram can be built from the
Feynman rules in momentum-space derived from ${\mathcal S}_{pp}$, ${\mathcal S}_{bulk}$ with propagators:
\be
\label{eq:props}
\ds P[W^a_p(t_a)W^b_{p'}(t_b)]&=&\ds \frac{1}{2} P^{aa}\delta_{ab}
\ds (2\pi)^d\delta^{(d)}(p+p'){\cal P}(p^2,t_a,t_b)\delta(t_a-t_b)\,,
\ee
where $P^{\phi\phi}=-\frac{1}{c_d}$, $P^{A_iA_j}=\delta_{ij}$, 
$P^{\sigma_{ij}\sigma_{kl}}=-\left(\delta_{ik}\delta_{jl}+\delta_{il}\delta_{jk}+(2-c_d)\delta_{ij}\delta_{kl}\right)$ 
and
\be
\label{eq:props_fac}
{\cal P}(p^2,t_a,t_b)=\frac{\text{i}}{p^2-\partial_{t_a}\partial_{t_b}}\,.
\ee

The terms involving time derivatives in the propagator must be Taylor
expanded and will end up hitting
the exponential factors $~e^{ip\cdot x}$ at source points, eventually generating extra factors of $v$ and its derivatives that has to be kept at the appropriate order.

The effective potential involves kinematic variables (positions, velocities,\ldots)
of the massive bodies at the {\it same time}, retardation effects dictated by GR
are reinstated in this framework by the knowledge of the trajectories via
their
derivatives at a point, and at any finite PN order only a finite number of trajectory derivatives
are necessary to reconstruct the effective action.

Note that working in the harmonic gauge simplifies considerably the computations since
linearized equations show that all polarizations satisfy D'Alembertian type
of equations, but this is a gauge artifact, analogous to what happens in
electromagnetism in the Feynman gauge, as gravity has only
2 propagating degrees of freedoms.
Indeed the resulting effective action will not be gauge-invariant, as
it is not observable, and as first noticed in
\cite{Schafer:1984mr} 
substituting into the Lagrangian the equations of motion is equivalent to a coordinate transformation.
As a consequence, adding to the
Lagrangian a term proportional to the equations of motion (henceforth, EOMs) does not alter the resulting
dynamics. 

On a technical point, we observe that the near-zone Lagrangian can contain terms involving
second or higher derivatives of the particle trajectory, leading to higher than second order
differential equations of motion. This problem can be solved by adding harmless terms proportional
to the square of the EOMs which neither alter the
resulting EOMs nor represent a change of coordinates,
but have the welcome feature of getting rid
of high derivative terms. Such a procedure was dubbed {\it double zero trick} \cite{Damour:1985mt} and roughly works as follows: taking for instance a
$a_1^2$ term at Newtonian order one can add a harmless term proportional to the square
of the equations of motion
\be
\label{eq:dz}
a_1^2\simeq a_1^2-\pa{\vec{a}_1+\frac{G m_2\vec{r}}{r^3}}^2=
-2\frac{G m_2 a_1^r}{r^3}-\frac{G^2 m_2^2}{r^4}\,,
\ee
(with notation $X^r\equiv \vec{X}\cdot\vec{r}/r$) without altering
the resulting EOMs.

In the following sections we report the results of Feynman diagrams at $G^n$ order with $n={3,4}$, involving integration
over $n-1$ internal momenta $k_1,\ldots,k_{n-1}$ and over the momentum $p$
exchanged between the two massive bodies.
While the integration diagrams in the $G$ and $G^2$ sub-sector can be readily performed via the use
of the formulae (\ref{eq:intp},\ref{eq:intkp}), at orders $G^3$ and  $G^4$ diagrams with up to $7$ propagators are present, and they can be computed with the technique of integration by part identities
\cite{Tkachov:1981wb,Chetyrkin:1981qh,Laporta:2001dd} enabling the reduction of all integrals to the master integrals.

The amplitudes related to any given Feynman diagram can be decomposed as
\begin{equation}
\begin{minipage}{2.5cm}
\begin{center}
\includegraphics[width=2.0cm]{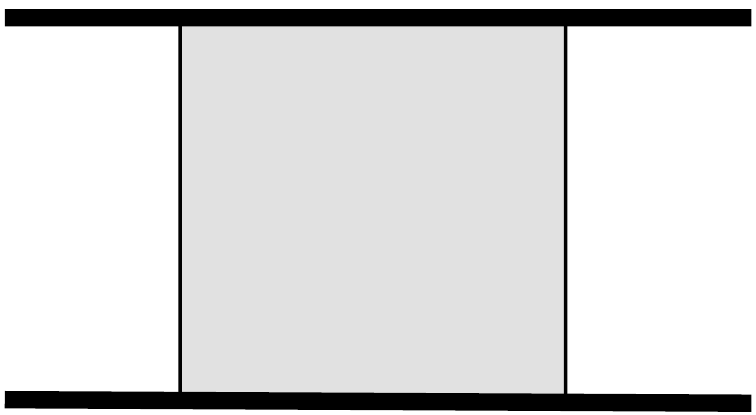}
\end{center}
\end{minipage}
\quad=\quad\sum c_a M_a\,,
\end{equation}
where the left hand side represents a generic EFT Feynman graph (thick horizontal lines represent the classical external sources which
exchange the dynamical fields $\phi$, $A$, $\sigma$ propagating in the shaded area), and the master integrals $M_a$ needed for our purpose are
\be
\label{eq:2loopMI}
\ba{rcl}
M^{(2loop)}_1(p)&\equiv&\ds\int \frac{d^dk_1}{(2\pi)^d}\frac{d^dk_2}{(2\pi)^d}\frac 1{k_1^2(p-k_2)^2(k_1-k_2)^2}\,,\\
M_2^{(2loop)}(p)&\equiv&\ds\int \frac{d^dk_1}{(2\pi)^d}\frac{d^dk_2}{(2\pi)^d}\frac 1{k_1^2(p-k_1)^2k_2^2(p-k_2)^2}\,,
\ea
\ee
at order $G^3$, and the additional
\be
\ba{rcl}
\label{eq:3loopMI}
M^{(3loop)}_1(p)&\equiv&\ds\int \frac{d^dk_1}{(2\pi)^d}\frac{d^dk_2}{(2\pi)^d}\frac{d^dk_3}{(2\pi)^d}
\frac 1{(p-k_3)^2k_2^2(k_1-k_2)^2(k_1-k_3)^2}\,,\\
M^{(3loop)}_2(p)&\equiv&\ds\int \frac{d^dk_1}{(2\pi)^d}\frac{d^dk_2}{(2\pi)^d}\frac{d^dk_3}{(2\pi)^d}
\frac 1{(p-k_2)^2(p-k_3)^2k_1^2k_3^2(k_1-k_2)^2}\,,\\
M^{(3loop)}_3(p)&\equiv&\ds\int \frac{d^dk_1}{(2\pi)^d}\frac{d^dk_2}{(2\pi)^d}\frac{d^dk_3}{(2\pi)^d}
\frac 1{(p-k_2)^2(p-k_3)^2k_1^2(k_1-k_2)^2(k_1-k_3)^2}\,,
\ea
\ee
at order $G^4$.

\begin{figure}[h]
\includegraphics[width=.4\linewidth]{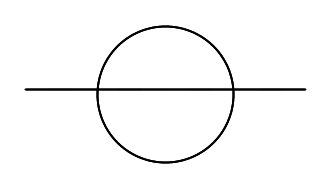}
\raisebox{0.73\height}{\includegraphics[width=.4\linewidth]{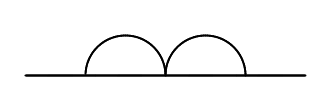}}
\caption{2-loop master integrals in eq.~(\ref{eq:2loopMI}).}
\label{fig:2lMI}
\end{figure}

\begin{figure}[h]
\includegraphics[width=.3\linewidth]{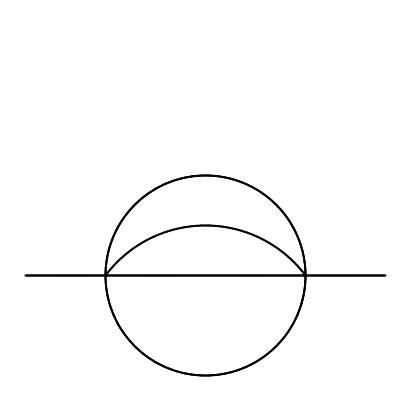}
\raisebox{0.28\height}{\includegraphics[width=.3\linewidth]{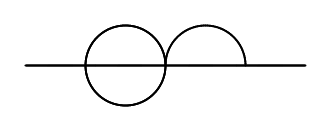}}
\raisebox{0.62\height}{\includegraphics[width=.3\linewidth]{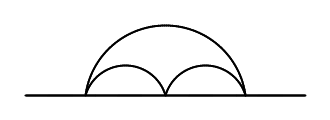}}
\caption{3-loop master integrals in eq.~(\ref{eq:3loopMI}).}
\label{fig:3lMI}
\end{figure}

Such master integrals can be represented in the language introduced in \cite{Foffa:2016rgu}, to which we remind the reader for further details, as
in figs.~\ref{fig:2lMI},\ref{fig:3lMI}.
The particle physics-oriented reader will appreciate the correspondence between the number of $d$-dimensional momentum integrations in the definitions of
the 
$M_a$s and the number of loops in their graphical representation, as well as between the denominator factors and the internal lines of the graphs.
The integrations by parts in multi-loop computations have been performed 
 by direct application of the integration by parts method
\cite{Tkachov:1981wb,Chetyrkin:1981qh} and independently via its
implementation in the Reduze code \cite{Studerus:2009ye}. Master integrals
have been computed by repeated application of equation
(\ref{eq:intkp}) reported in the appendix,
and the Lagrangian contribution is finally obtained by taking the Fourier transform with respect to $p$,
which is the momentum carried by the external legs in figs.~\ref{fig:2lMI},\ref{fig:3lMI}.

For the actual computation we use the Feyncalc software
\cite{Shtabovenko:2016olh} in Mathematica.

\section{Results of diagram computation}
\label{sec:results}
The result for the $G$ and $G^2$ near zone computations are already published
in eq.(13,14,17,18,19) of \cite{Foffa:2012rn}, with the list of relative
diagrams in fig.~1 and 2 of the same paper (respectively 3 and 23 diagrams),
and will not be reported here.

\subsection{The 4PN ${\bm G^3}$ sector}
The $G^3$ sector receives contribution from 212 diagrams which are available,
together with their values, at {\tt http://fiteoweb.unige.ch/$\sim$foffa/research/research.html}.
Their sum gives \footnote{We denote the scalar product among two vectors
$\vec a$, $\vec b$ as $a.b$, define $\vec v\equiv \vec v_1-\vec v_2$ and an
upper $r$ stands for scalar product with unit vector separation between the two
bodies, e.g. $v^r\equiv (v_1-v_2).(x_1-x_2)/r$.}:
\allowdisplaybreaks
\be
\label{eq:4PNG3graphs}
&&{\cal L}^{(4PN\,graphs)}_{G^3}=\frac{G^3 m_1^3 m_2}{r^3}\left[v_2.a_1\left(\frac{4859}{120}v_1^r
-\frac{12187}{360}v_2^r\right) r+a_1^r\left(\frac{493}{18}v_1^2+\frac{5143}{720}\left(v_2^2-2v_1.v_2\right)\right.\right.\nonumber\\
&&+\left.\frac{6347}{240}{v_2^r}^2-\frac{11497}{120}v_1^r v_2^r\right) r
-\frac{12187}{720}r a_2^rv_1^2+\frac{15787}{720}v_1^4+\frac{19927}{720}v_1^2 \left(v_2^2-2v_1.v_2\right)+\frac{43}{2}v_1.v_2 v.v_2\nonumber\\
&&+\frac{91}{16}v_2^4+v_1^2\left(\frac{15487}{120}v_1^r v_2^r-\frac{3373}{30}{v_1^r}^2
-\frac{16387}{240}{v_2^r}^2\right)+v_1.v_2\left(7v_1^r v_2^r+\frac{11557}{240} {v_1^r}^2-2{v_2^r}^2\right)\nonumber\\
&&\left.+v_2^2\left(\frac{7}{4}{v_2^r}^2-\frac{1}{8}{v_1^r}^2-\frac{7}{2}v_1^r v_2^r\right)
+\!\left(\frac{1661}{18}{v_1^r}^2-\frac{12613}{144}v_1^r v_2^r-\frac{15}{4}{v_2^r}^2\right) {v_1^r}^2\!\right]\nonumber\\
&&+\frac{G^3 m_1^2 m_2^2}{r^3}\left\{a_1^r\left[\left(\frac{20129}{450}-\frac{21\pi^2}{32}\right)v_1^2
+\left(\frac{23297}{600}-\frac{43}{128}\pi^2\right)\left(v_2^2-2v_1.v_2\right)+\left(\frac{123\pi^2}{128}-\frac{431}{24}\right){v_2^r}^2\right]r\right.\nonumber\\
&&+r v_2.a_1\left(\frac{425}{36}-\frac{41\pi^2}{64}\right)v_1^r+\left(\frac{3953}{90}+\frac{133\pi^2}{1024}\right)v_1^4
+\left(\frac{31759}{720}+\frac{133\pi^2}{1024}\right)\left(v_1^2 v_2^2+2(v_1.v_2)^2-4v_1^2 v_1.v_2\right)\nonumber\\
&&+v_1.v_2\left(\frac{90223}{300}-\frac{447\pi^2}{256}\right)v^r v_1^r- 2v_1.v_2 v_1^r v_2^r+v_1^2\left[\left(\frac{447\pi^2}{512}-\frac{22687}{150}\right){v_1^r}(v_1^r-2v_2^r)\right.\nonumber\\
&&+\left.\left(\frac{447\pi^2}{512}-\frac{90223}{600}\right){v_2^r}^2\right]+\left[\left(\frac{2155\pi^2}{256}-\frac{7348}{15}\right)v_1^r v_2^r
+\left(\frac{1837}{15}-\frac{2155\pi^2}{1024}\right){v_1^r}^2\right.\nonumber\\
&&+\left.\left.\left(\frac{29617}{80}-\frac{6465\pi^2}{1024}\right){v_2^r}^2\right] {v_1^r}^2\right\}+{\cal L}^{(4PNa^2)}_{G^3fin}+{\cal L}^{(4PN)}_{G^3pole}\nonumber\\
&&+\log{\bar r}\left\{
\frac{G^3 m_1^3 m_2}{r^3}\left[v_2.a_1\left(37v_2^r-48v_1^r\right) r+r a_1^r\left(16 v_1.v_2-32v_1^2
-8 v_2^2-\frac{51}{2}{v_2^r}^2+84v_1^r v_2^r\right) \right.\right.\nonumber\\
&&\left.+\frac{37}{2}v_1^2\left(r a_2^r -v^2\right)+v_1^2\left(\frac{195}{2}{v_1^r}^2-111v_1^r v_2^r
+\frac{111}{2}{v_2^r}^2\right)-42v_1.v_2 {v_1^r}^2-70{v_1^r}^3v^r\!\right]\nonumber\\
&&+\left.\frac{G^3 m_1^2 m_2^2}{r^3}\left[14 r a_1^rv^2+42(v_1^2 v^r
-2v_1.v_2 v_1^r)v^r-70{v_1^r}^2\left({v_1^r}^2-4v_1^r v_2^r+3{v_2^r}^2\right)\!\right]\right\}\,,
\ee
with $\bar{r}\equiv r/L_0$, ${\cal L}^{(4PNa^2)}_{G^3fin}$ gathering the non-singular contributions quadratic in accelerations or higher derivatives
\be
\label{eq:4PNG3findz}
{\cal L}^{(4PNa^2)}_{G^3fin}&=&\frac{G^3 m_1^2 m_2^2}{r}\left[\left(\frac{656}{25}-\frac{225}{64}\pi^2\right)a_1^2-\left(\frac{46657}{3600}+\frac{309}{128}\pi^2\right)a_1.a_2
+\left(\frac{1547}{225}+\frac{15}{64}\pi^2\right){a_1^r}^2\right.\nonumber\\
&&\left.+\left(\frac{38917}{3600}+\frac{25}{128}\pi^2\right) a_1^r a_2^r\right]+\frac{G^3 m_1^3 m_2}{r}\left[\frac{2131}{150}{a_1^r}^2+\frac{3407}{720} a_1^r a_2^r-\frac{4891}{450}a_1^2-\frac{8029}{240}a_1.a_2\right]\nonumber\\
&&-3(d-3){\cal L}^{(4PNa^2)}_{G^3pole}\log{\bar r}\,,
\ee
and ${\cal L}^{(4PN)}_{G^3pole}$ gathering the singular part
\be
\label{eq:4PNG3pole}
\ds{\cal L}^{(4PN)}_{G^3pole}&=&\ds\frac{G^3 m_1^3 m_2}{(d-3)r^3}\left[r a_1^r\left(\frac{11}{6}v_1^2-\frac{11}{2}{v_2^r}^2+\frac{8}{3}v^2+14 {v^r}^2\right)-\frac{37}{3}v.a_1 r v^r
+\frac{11}{3}v_2.a_1 r v_1^r\right]\nonumber\\
&&-\frac{14}{3}\frac{G^3 m_1^2 m_2^2}{(d-3)r^2}a_1^r \left(v^2-3{v^r}^2\right)+{\cal L}^{(4PNa^2)}_{G^3pole}
\ee
in which we have separated the part
quadratic in acceleration or higher derivative terms
\be\label{eq:4PNG3poleHD}
\ds{\cal L}^{(4PNa^2)}_{G^3pole}&=&\ds\frac{G^3 m_1^3 m_2}{(d-3)r^3}\left[
\left(\frac{133}{5}a_1^2-\frac{11}{2}a_1.a_2+\frac{16}{5}{a_1^r}^2+\frac{17}{6}a_1^r a_2^r\right)r^2\right]
\nonumber\\
&&+\frac{G^3 m_1^2 m_2^2}{(d-3)r^3}\left[\left(\frac{10}{3}a_1^2+\frac{124}{15}a_1.a_2+\frac{10}{3}{a_1^r}^2+\frac{68}{15}a_1^r a_2^r\right)r^2\right]\,.
\ee

\subsection{The 4PN ${\bm G^4}$ and ${\bm G^5}$ sectors}
The $G^4$ sector receives contribution from 317 diagrams, also available with
their
individual values at {\tt http://fiteoweb.unige.ch/$\sim$foffa/research/research.html} (see also \cite{Cristofoli:2018bex} for a detailed sample calculation) whose sum gives
\be\label{4PNG4graphs}
&&{\cal L}^{(4PN\,graphs)}_{G^4}=\frac{G^4 m_1^4 m_2}{r^4}\left[\frac{50}{9}v_1^2-\frac{227}{36}v_1.v_2+\frac{15}{16}v_2^2-\frac{422}{9}{v_1^r}^2+\frac{817}{18}v_1^r v_2^r+\frac{9}{4}{v_2^r}^2\right]\nonumber\\
&&+\frac{G^4 m_1^3 m_2^2}{r^4}\left[\left(\frac{466}{9}-\frac{157}{32}\pi^2\right)v_1^2+\left(\frac{69}{16}\pi^2-\frac{1201}{18}\right)v_1.v_2+\left(\frac{19}{32}\pi^2+\frac{323}{18}\right)v_2^2\right.\nonumber\\
&&\left.\left(\frac{2363}{96}\pi^2-\frac{1468}{9}\right){v_1^r}^2+\left(\frac{1958}{9}-\frac{1307}{48}\pi^2\right)v_1^r v_2^r+\left(\frac{251}{96}\pi^2-\frac{382}{9}\right){v_2^r}^2\right]\nonumber\\
&&+\left(\frac{1}{(d-3)}-4 \log{\bar r}\right)\left\{\frac{G^4 m_1^3 m_2^2}{3r^4}\left[v_1^2-13 v_1.v_2+12 v_2^2+44 v_1^r v^r\right]\right.\nonumber\\
&&\left.+\frac{23}{3}\frac{G^4 m_1^4 m_2}{r^4}\left[v.v_1-4v_1^r v^r\right]\right\}.
\ee
For later convenience, we recall here the also sum of the 50 $G^5$ diagrams in fig.1 of
\cite{Foffa:2016rgu}:
\be
\label{eq:4PNG5graphs}
&&{\cal L}^{(4PN\,graphs)}_{G^5}=\frac{3}{8}\frac{G^5 m_1^5 m_2}{r^5}+\frac{31}{3}\frac{G^5 m_1^4 m_2^2}{r^5}+\frac{141}{8}\frac{G^5 m_1^3 m_2^3}{r^5}\,.
\ee

\subsection{The 3PN sector}
Divergences occur in the effective Lagrangian starting from 3PN order, 
which have to be taken into account along with
the 4PN $G^2$ \cite{Foffa:2012rn} , $G^3$ and $G^4$ ones.

We consider here the {\it bare} 3PN Lagrangian derived from near-zone
graphs computed in \cite{Foffa:2011ub},
without the use of {\it coordinate shift} performed in \cite{Blanchet:2003gy}
 and \cite{Foffa:2011ub},
(but we add double zeroes to the finite part to get rid of terms quadratic in accelerations or
higher derivative) with the result that our regularized 3PN
Lagrangian coincides with the one of
\cite{Blanchet:2003gy} only up to $G^2$ terms included. In total we have:
\be
{\cal L}^{(3PN)}={\cal L}^{(3PN)}_{G+G^2}+ {\cal L}^{(3PN)}_{G^3+G^4}+{\cal L}^{(3PN)}_{pole}
\ee
with
\be
\label{eq:3PN}
&&{\cal L}^{(3PN)}_{G^3+G^4}=\frac{G^3 m_1^3 m_2}{r^3}\left[\frac{55}{9}v_1^2
-\frac{137}{18}v_1.v_2+\frac{5}{4}v_2^2-\frac{67}{4}{v_1^r}^2+\frac{29}{2}v_1^r.v_2^r+\frac{3}{2}{v_2^r}^2\right]\nonumber\\
&&+\frac{G^3 m_1^2 m_2^2}{r^3}\left[-\frac{305}{72}v_1^2+\frac{439}{144}v_1.v_2+\frac{383}{24}{v_1^r}^2
-\frac{889}{48}v_1^r.v_2^r+\frac{41\pi^2}{64}\left(v.v_1-3v^r v_1^r\right)\right]\nonumber\\
&&-\frac{3}{8}\frac{G^4 m_1^4 m_2}{r^4}-\frac{67}{3}\frac{G^4 m_1^3 m_2^2}{r^4}+\frac{11}{3}\log{\bar r}\left[-\frac{G^3 m_1^3 m_2}{r^3}\left(v.v_1-3 v^r v_1^r\right)+2\frac{G^4 m_1^3 m_2^2}{r^4}\right]\,,
\ee
and
\be\label{3PNpole}
{\cal L}^{(3PN)}_{pole}=-\frac{11G^2 m_1^2 m_2}{3(d-3)}\left(a_1^2+2 a_1.a_2\right)+\frac{11G^3 m_1^3 m_2}{3(d-3)r^2}a_1^r\,,
\ee
while ${\cal L}^{(3PN)}_{G+G^2}$ is the same as reported in sec. 7.2 of \cite{Blanchet:2013haa}.

\section{The regularized 3PN and 4PN Lagrangian}
\label{sec:reg_lag}
We now put together the results of the previous section, transforming eq. (\ref{eq:4PNG3findz}) by means of the double zero trick, while leaving
unaltered analogous higher derivative terms contained in the divergent
pieces, given by eq.(\ref{eq:4PNG3poleHD}).
We also include here the 4PN terms generated by applying the double zero
trick to the 2PN, 3PN, 4PN $G$, $G^2$ and $G^3$ finite terms, according to the equations reported in Appendix \ref{app:dztricks}.

As we mentioned in the introduction, for the sake of comparison with previous results we 
add by hand the local-in-time contribution from the tail term ${\cal L}_{tail}$.
The complete expression of the dimensionally regularized ${\cal L}_{tail}$,
inclusive of its non-local-in-time part obtained earlier in \cite{Blanchet:1987wq}, has first been computed in
\cite{Foffa:2011np,Galley:2015kus}, and re-derived in \cite{Bernard:2017bvn}
\be
\label{eq:tailQQ}
{\cal L}_{tail}&\equiv&-\frac{G^2 M}5\left\{\int_t \left[\frac{1}{d-3}-\frac{41}{30}-2\log{\bar{r}}\right]\dddot{\cal Q}_{ij}^2(t)+\int_{k_0} 2 k_0^6
\log{\left(2e^{\gamma_E}k_0 r\right)}\left|{\cal Q}^{ij}(k_0)\right|^2\right\}\,,
\nonumber\\
\ee
which takes the explicit form (\ref{eq:QQ}) after the
quadrupole ${\cal Q}_{ij}$ is expressed in terms of orbital variables
\footnote{The part involving $\gamma_E$ could well be incorporated into the
local-in-time part, here we decide it to keep into the non-local part.}.

We will retain the local-in-time part, including the pole, which we need
to compare our results with those presented in \cite{Damour:2014jta} and 
\cite{Marchand:2017pir}.
We discuss the proper incorporation of the tail terms as well as the
renormalization of the effective theory in \cite{Foffa:2019yfl},
to which we refer the reader for details. 

In what follows we show that the regularized local-in-time Lagrangian
obtained with this procedure agrees with previous results in the literature.
The sum of all the terms discussed so far can be written as
\be\label{Lreg}
{\cal L}^{(3+4PN)}_{reg}={\cal L}^{(3PN)}_{fin} + {\cal L}^{(4PN)}_{fin}+\frac 1{d-3}{\cal L}_{poles}\,.
\ee
The quantity $\frac 1{d-3}{\cal L}_{poles}$ beside all divergent terms (including the tail ones) includes some {\it ad-hoc} finite
pieces to ensure that it vanishes on-shell; such finite terms,
which are included in eq. (\ref{sec:shift_to_0}) \footnote{These
finite terms can also be read in eqs. (3.14,3.16,3.17,4.19) of \cite{Foffa:2019yfl}.},
have been consequently added with opposite sign into ${\cal L}^{(3PN)}_{fin} + {\cal L}^{(4PN)}_{fin}$ :
the result has been organized in this way for ease of
comparison with the literature, and we are now going to analyze the various
contributions separately.

\subsection{${\cal L}_{poles}$}
The term containing ${\cal L}_{poles}$ has both UV and IR divergences.
The former are expected to warn about failure to describe short
distance features of the model, however for non-spinning bodies
short-distance departures from the point particle limit enter only at
5PN order, as demonstrated by the {\it effacement principle}
\cite{Damour:1982wm}, and all UV divergences met in the present work
must drop out of physical observables.

On the other hand the artificial splitting in near and far zone
involving gravitational modes with hard and soft momenta
introduces in the near-zone computation spurious IR divergences,
some of which are contributing to ${\cal L}_{poles}$,
while the remaining IR divergences are canceled by the addition of
${\cal L}_{tail}$.

However all the terms lumped together in ${\cal L}_{poles}$ do not affect
physical observables (like the energy-frequency relation, or the
periastron advance) because, as can be see from their explicit form
reported in (\ref{sec:shift_to_0}), ${\cal L}_{poles}$ is proportional to 
the EOMs, including ${\cal O}(d-3)$ terms and 1PN corrections
to the leading order EOM when necessary.

In the Fokker action treatment of \cite{Bernard:2015njp} this feature
has been exploited to remove such poles by means of a coordinate shift.
As ${\cal L}_{poles}$ contains both UV and IR divergences,
removing them via a coordinate shift, although providing the correct result,
is not well motivated from the point of view of field theory, where UV
divergences should be removed by counterterms and all IR divergences should be
treated together, irrespectively of the fact that they vanish on-shell
(like the ones in ${\cal L}_{poles}$) or not (like the one canceled by ${\cal L}_{tail}$).
A fully field-theory-motivated, as well as self contained and ambiguity-free
treatment of such matter, is beyond the scope of this
work and is the subject of \cite{Foffa:2019yfl}.

\subsection{${\cal L}^{(3PN)}_{fin} + {\cal L}^{(4PN)}_{fin}$}
\label{sec:fin}
The two finite terms ${\cal L}^{(3PN)}_{fin}$ and ${\cal L}^{(4PN)}_{fin}$
include
\begin{itemize}
\item the finite part of the result of diagram computations
\item the addition of double zeroes to get rid of higher derivatives
\item the finite part of the instantaneous tail term
\item the opposite of finite pieces appearing in ${\cal L}_{poles}/(d-3)$.
\end{itemize}
For the 3PN part one gets
\be\label{3PNfin}
{\cal L}^{(3PN)}_{fin}
&=&{\cal L}^{(3PN)}_{G+G^2}+\frac{G^3 m_1^3 m_2}{r^3}\left[\frac{209}{18}v_1^2
-\frac{118}{9}v_1.v_2+\frac{5}{4}v_2^2-\frac{443}{12}{v_1^r}^2+\frac{104}{3}v_1^r.v_2^r+\frac{3}{2}{v_2^r}^2\right]\nonumber\\
&&+\frac{G^3 m_1^2 m_2^2}{r^3}\left[-\frac{305}{72}v_1^2+\frac{439}{144}v_1.v_2+\frac{383}{24}{v_1^r}^2
-\frac{889}{48}v_1^r.v_2^r+\frac{41\pi^2}{64}\left(v.v_1-3v^r v_1^r\right)\right]\nonumber\\
&&-\frac{3}{8}\frac{G^4 m_1^4 m_2}{r^4}-\frac{67}{3}\frac{G^4 m_1^3 m_2^2}{r^4}+\frac{22}{3}\log{\bar r}\left[-\frac{G^3 m_1^3 m_2}{r^3}\left(v.v_1-3 v^r v_1^r\right)+\frac{G^4 m_1^3 m_2^2}{r^4}\right]\,.
\ee

As to the 4PN part, it can be decomposed into
\be
{\cal L}^{(4PN)}_{fin}=\frac 7{256}m_1v_1^{10}+{\cal L}^{(4PN)}_G+{\cal L}^{(4PN)}_{G^2}+{\cal L}^{(4PN)}_{G^3}+
{\cal L}^{(4PN)}_{G^4}+{\cal L}^{(4PN)}_{G^5}\,,
\ee
where the finite terms ${\cal L}^{(4PN)}_G+{\cal L}^{(4PN)}_{G^2}$ are have already been computed and published, see eqs.(13,26) in \cite{Foffa:2012rn},  and the others read:
\be\label{4PNG3reg}
&&{\cal L}^{(4PN)}_{G^3}=\frac{G^3 m_1^3 m_2}{r^3}\left[v_2.a_1\left(\frac{3763}{240}v_2^r
-\frac{18719}{720}v_1^r\right) r+a_1^r\left(-\frac{18719}{1440}v_1^2-\frac{95119}{7200}v^2+\frac{1309}{48}{v_2^r}^2-\frac{75}{4}v_1^r v_2^r\right) r\right.\nonumber\\
&&+\frac{3763}{480}r a_2^r v_1^2-\frac{231}{160}v_1^4+\frac{1397}{480}v_1^2 v_2^2 -\frac{433}{60}v_1^2 v_1.v_2
+\frac{43}{2}v_1.v_2 v.v_2+\frac{91}{16}v_2^4\nonumber\\
&&+v_1^2\left(\frac{15349}{480}{v_1^r}^2-\frac{4381}{60}v_1^r v_2^r
+\frac{16729}{480}{v_2^r}^2\right)+v_1.v_2\left(7v_1^r v_2^r+\frac{43}{16} {v_1^r}^2-2{v_2^r}^2\right)\nonumber\\
&&\left.+v_2^2\left(\frac{7}{4}{v_2^r}^2-\frac{1}{8}{v_1^r}^2-\frac{7}{2}v_1^r v_2^r\right)
+\!\left(\frac{43}{6}{v_1^r}^2-\frac{119}{48}v_1^r v_2^r-\frac{15}{4}{v_2^r}^2\right) {v_1^r}^2\!\right]\nonumber\\
&&+\frac{G^3 m_1^2 m_2^2}{r^3}\left\{a_1^r\left[\left(\frac{349207}{7200}-\frac{43}{128}\pi^2\right)v^2+\left(\frac{123\pi^2}{128}-\frac{2005}{96}\right){v_2^r}^2\right]\!\!r\right.\nonumber\\
&&+r (2 v_1^r v_2.a_1+ a_1^r v_1^2)\left(\frac{1099}{288}-\frac{41\pi^2}{128}\right)+\frac{383}{192}v_1^4+\left(\frac{21427}{480}+\frac{133\pi^2}{1024}\right)\left(v_1^2 v^2-2v_1.v_2 v.v_1\right)\nonumber\\
&&-\frac{55}{24}v_1^2 v_1.v_2+v_1.v_2\left(\frac{31687}{150}-\frac{447\pi^2}{256}\right)v^r v_1^r- 4v_1.v_2 v_1^r v_2^r\nonumber\\
&&+v_1^2\left[\frac{260921}{1200}v_1^r v_2^r-\frac{265721}{2400}{v_1^r}^2-\frac{62399}{600}{v_2^r}^2+\frac{447\pi^2}{512}{v^r}^2\right]\nonumber\\
&&\left.+\left[\frac{155563}{2880}{v_1^r}^2-\frac{155593}{720}v_1^r v_2^r+\frac{78719}{480}{v_2^r}^2-\frac{2155\pi^2}{1024}\left({v_1^r}^2-4v_1^r v_2^r+3{v_2^r}^2\right)\right] {v_1^r}^2\right\}\nonumber\\
&&+\log{\bar r}\left\{
\frac{G^3 m_1^3 m_2}{r^3}\left[v_2.a_1\left(28v_1^r-\frac{62}{3}v_2^r\right) r+r a_1^r\left(14v_1^2+\frac{338}{15}v^2-25{v_2^r}^2+28v_1^r v_2^r\right) \right.\right.\nonumber\\
&&\left.-\frac{31}{3}v_1^2\left(r a_2^r -v^2\right)+v_1^2\left(14{v_1^r}^2-31{v^r}^2\right)-14v_1.v_2 {v_1^r}^2-\frac{70}{3}{v_1^r}^3v^r\!\right]\nonumber\\
&&\left.+\frac{G^3 m_1^2 m_2^2}{r^3}\left[\frac{34}{15} r a_1^r v^2+4(v_1^2 v^2-2 v_1.v_2 v.v_1)-\frac{62}{5}(v_1^2 v^r 
-2v_1.v_2 v_1^r)v^r+\frac{2}{3}{v_1^r}^2\left({v_1^r}^2-4v_1^r v_2^r+3{v_2^r}^2\right)\!\right]\right\}\,.\nonumber\\
\ee
\be\label{4PNG4reg}
&&{\cal L}^{(4PN)}_{G^4}=\frac{G^4 m_1^4 m_2}{r^4}\left[-\frac{177749}{3600}v_1^2+\frac{175049}{3600}v_1.v_2+\frac{15}{16}v_2^2+\frac{242309}{900}{v_1^r}^2-\frac{243659}{900}v_1^r v_2^r+\frac{9}{4}{v_2^r}^2\right]\nonumber\\
&&+\frac{G^4 m_1^3 m_2^2}{r^4}\left[-\left(\frac{91369}{7200}+\frac{15}{32}\pi^2\right)v_1^2+\left(\frac{103}{16}\pi^2-\frac{18083}{240}\right)v_1.v_2+\left(\frac{714259}{7200}-\frac{191}{32}\pi^2\right)v_2^2\right.\nonumber\\
&&\left.\left(\frac{659}{96}\pi^2-\frac{349369}{7200}\right){v_1^r}^2+\left(\frac{430573}{720}-\frac{1715}{48}\pi^2\right)v_1^r v_2^r+\left(\frac{2771}{96}\pi^2-\frac{414029}{800}\right){v_2^r}^2\right]\nonumber\\
&&+\log{\bar r}\left\{\frac{G^4 m_1^3 m_2^2}{r^4}\left[8v_1^2+\frac{181}{15}v_1.v_2-\frac{356}{15} v_2^2+\frac{118}{5} {v_1^r}^2-\frac{944}{5} v_1^r v_2^r+\frac{2258}{15} {v_2^r}^2\right]\right.\nonumber\\
&&\left.+\frac{986}{15}\frac{G^4 m_1^4 m_2}{r^4}\left[v.v_1-4v_1^r v^r\right]\right\}\,,
\ee
\be\label{4PNG5reg}
{\cal L}^{(4PN)}_{G^5}&=&\frac{3}{8}\frac{G^5 m_1^5 m_2}{r^5}+\frac{G^5 m_1^4 m_2^2}{r^5}\left(\frac{88331}{3600}+\frac{105}{32}\pi^2-\frac{216}{5}\log{\bar r}\right)\nonumber\\
&&+\frac{G^5 m_1^3 m_2^3}{r^5}\left(\frac{73813}{800}-\frac{71}{32}\pi^2-\frac{83}{15}\log{\bar r}\right).
\ee

${\cal L}^{(3PN)}_{fin}+{\cal L}^{(4PN)}_{fin}$  can be compared directly with the result published in app.~A of \cite{Bernard:2017ktp} and we find that they differ by a term proportional to the equations of motion:
\be
\Delta{\cal L}=\vec{Z}_1\cdot\vec{Eq}^0_1\,,
\ee
with
\be
\vec{Z}_1&=&-\frac{G^3 m_1^2 m_2}{r^3}\left(\left\{m_1\left[\frac{109}{140}\left(1+\frac12 v_1^2-\frac32{v_2^r}^2\right)-\frac{8317}{600}v^2+\frac{31213}{600}{v^r}^2+\left(\frac{8}{15}v^2-14 {v^r}^2\right)\log{\bar{r}}\right]\right.\right.\nonumber\\
&&+m_2\left[\frac{82391}{1200}{v^r}^2-\frac{5679}{200}v^2+\left(\frac{2}{5}{v^r}^2-\frac{26}{15}v^2\right)\log{\bar{r}}\right]+\frac{G}{r}\left[\left(\frac{38}{5}m_1^2+\frac{467}{15}m_1m_2+\frac{688}{15}m_2^2\right)\log{\bar{r}}\right.\nonumber\\
&&-\left.\left.\left(\frac{93}{10}m_1^2+\frac{1591}{140}m_1m_2+\frac{1307}{42}m_2^2\right)\right]\right\}\vec{r}\nonumber\\
&&\left.+\frac{109}{140}m_1 v_1^r r\vec{v}_2+v^r\left[m_1\left(16 \log{\bar{r}}-\frac{59921}{1680}\right)-m_2\left(8 \log{\bar{r}}+\frac{2293}{80}\right)\right]r\vec{v}\right)\,,
\ee
and $\vec{Eq}^0_1$ is defined in (\ref{eq:eq0}).

We have thus proven that the regularized Lagrangian derived within the EFT framework is fully consistent with the results obtained by
other groups; ${\cal L}^{3PN}_{fin}+{\cal L}^{4PN}_{fin}$ can thus {\it de facto} be used as the finite Lagrangian in that it provides
the correct dynamics for physical observables, like energy of circular orbits and periastron precession.

\section{Conclusions}
\label{sec:conclusions}
We have derived the last missing ingredients and completed the calculation
of the regularized conservative, local-in-time Lagrangian at fourth post-Newtonian
order with effective field theory methods.

We have also shown that our results for the $G^3$ and $G^4$
sectors are consistent with the literature
since following the manipulations presented in \cite{Marchand:2017pir} we arrive at the same answer, also consistent with \cite{Damour:2014jta}.
However, we remark that the treatments presently available in literature, while providing the correct physical
answer, still present delicate aspects related to the need of external inputs like
the matching to self-force computations as in the ADM
approach \cite{Bini:2013zaa},
or to the use of a coordinate shift to remove both
UV and (some of the) IR divergences, as in the Fokker action one \cite{Bernard:2015njp}.
The companion paper \cite{Foffa:2019yfl} is devoted to the
concurrent resolution/clarification of both these issues.

\section*{Acknowledgments}
The authors wish to thank Rafael Porto and Ira Rothstein for fruitful exchanges, which eventually led to the project reported in \cite{Foffa:2019yfl}.
We wish also to thank ICTP-SAIFR, supported by FAPESP grant 2016/01343-7, for the organization of the workshop ``Analytic methods in General Relativity", where many stimulating discussions took place, as well as
the Mainz Institute for Theoretical Physics (MITP) for its hospitality and its partial support during the completion of this work.
The work of RS has been supported for part of the duration of the present work by the FAPESP grant n. 2012/14132-3 and wishes to thank the University of Geneva Physics Department for hospitality and support during his frequent visits.
RS acknowledges the High Performance Computing Center at UFRN. SF is supported by the Fonds National Suisse and by the SwissMap NCCR and thanks the International Institute of Physics
of UFRN in Natal for kind hospitality and support during his visit.

\appendix
\section{Fundamental integrals}
\label{app:int}
\be
\label{eq:intp}
\int \frac{d^dp}{(2\pi)^d}\frac{e^{i\Pp.\X}}{p^{2a}}=
2^{-2a}\pi^{-d/2}\frac{\Gamma(d/2-a)}{\Gamma(a)}r^{2a-d}\,,
\ee
\be
\label{eq:intkp}
\int \frac{d^dk}{(2\pi)^d}\frac 1{\pa{k-p}^{2a}k^{2b}}=\frac 1{(4\pi)^{d/2}}
\frac{\Gamma(d/2-a)\Gamma(d/2-b)\Gamma(a+b-d/2)}{\Gamma(a)\Gamma(b)\Gamma(d-a-b)}\,.
\ee

\section{Tail terms}
In terms of relative variables the instantaneous part (that is, the
terms not involving $k_0$) of eq.(\ref{eq:tailQQ}) can be rewritten as
\be
\label{eq:QQ}
\ds{\cal L}_{tail,inst}&\equiv&\ds-\frac{2G^2 m_1^2 m_2^2}{5(d-3)M}\left[ r^2 {\dot a}^2+\frac{(r.\dot{a})^2}{3}+6 a.\dot{a}\ r.v +6 v.\dot{a} r.a-4 r.\dot{a} v.a +9 a^2 v^2 +3 (v.a)^2\right]\nonumber\\
&&\ds+\frac{G^4 m_1^3 m_2^2}{r^4}
\pa{\frac{656}{75}v^2-\frac{608}{75}{v^r}^2}+\frac{G^4 m_1^3 m_2^2}{r^4}
\left(\frac{64}{5}v^2-\frac{176}{15}{v^r}^2\right)\log{\bar{r}}\,.
\ee
The non-instantaneous piece becomes also instantaneous in the case of
circular orbit and its contribution is necessary to recover the correct
logarithmic part of the well known result of the energy of circular
orbits.

\section{Explicit expression of ${\cal L}_{poles}$}
The equations of motion up to order $\varepsilon\equiv d-3$ are
\be
\vec{Eq}_{1,2}=\vec{Eq}_{1,2}^0+\varepsilon \vec{Eq}_{1,2}^\varepsilon
\ee
with
\be
\label{eq:eq0}
\vec{Eq}_1^0=\vec{a}_1+\frac{G m_2}{r^3}\left[1-4 v_1.v_2+v_1^2+2 v_2^2-\frac32 {v_2^r}^2\right]\vec{r}-\frac{G^2 m_2}{r^4}(5m_1+4m_2)\vec{r}-\frac{G m_2}{r^2}(4v_1^r-3v_2^r)\vec{v}\nonumber\\
\ee
\be
\ba{rcl}
\vec{Eq}_1^\varepsilon&=&\ds\frac{G_N m_2}{2r^3}\biggl\{
\bigl[8\pa{\log{\bar{r}}-1}v_1^r+(5-6\log{\bar{r}})v_2^r\bigr]\vec{v}+\biggr.\\
&&\ds\left[\left(3-2\log{\bar{r}}\right)\left(1-v_1^2\right)+
4v^2\pa{1-\log{\bar{r}}}+{v_2^r}^2\pa{3\log{\bar{r}}-\frac{11}2}\right.\\
&&\ds\left.\left.
+\frac{G_Nm_2}r4\pa{4\log{\bar{r}}-3}+\frac{G_Nm_1}r\pa{20\log{\bar{r}}-17}
\right]\vec{r}\right\}\,,
\ea
\ee
and analogue expressions for $\vec{Eq}_{2}^{0,\varepsilon}$.

If we define
\be
\vec{K}_{1,2}\equiv \frac12\left(\vec{Eq}_{1,2}^0-\varepsilon \vec{Eq}_{1,2}^\varepsilon\right)
\ee
one can rewrite $\hat{\cal L}^{poles}$ as a term proportional to the equations of motion
at ${\cal O}(\varepsilon)$, which is the needed order as  ${\cal L}^{poles}$ appears with a prefactor $\varepsilon^{-1}$ in eq.(\ref{Lreg})
\be\label{sec:shift_to_0}
{\cal L}^{poles}&=&G^2 m_1^2\left\{-\frac{11}{3} \left[M+\frac{3}{2} m_1 v_1^2+m_2\left(\frac32 v_1^2+\frac{1}{2}v^2\right)\right]-\frac{19}{15} m_2r  a_2^r\right.\nonumber\\
&+&\left.\frac{7G m_1 m_2}{r}+\frac{19 G m_2^2}{15r}
+\frac{m_2^2}{M}\left(\frac{12}{5}r a^r-\frac{18}{5}v^2\right)\right\} 2 Eq_1.K_1\nonumber\\
&+&G^2 m_1^2 m_2\left[-\frac{22}{3}+\frac{16}{5}v_1.v_2-\frac{134}{15}v_1^2-\frac{79}{15}v_2^2+\frac{34}{15}r a_2^r-\frac{14}{3}r a_1^r+\frac{379G m_2}{30r}+\frac{23 G m_1^2}{5Mr}\right.\nonumber\\
&+&\left.\frac{m_2}{M}\left(\frac{18}{5}v^2-\frac{12}{5}r a^r\right)\right]\left(Eq_1.K_2+Eq_2.K_1\right)\nonumber\\
&+&\frac{G^3 m_1^2 m_2}{r^3}\left[\left(\frac{206}{15}m_1+\frac{66}{5}m_2\right) r.K_1 r.Eq_1+\left(\frac{89}{15}m_1+\frac{31}{5}m_2\right)\left(r.K_2 r.Eq_1+r.K_1 r.Eq_2\right)\right]\nonumber\\
&+&G^2 m_1^2m_2\left[\frac{12 m_2}{5 M} v.K_1 v.Eq_1-\left(\frac{64}{5}+\frac{6 m_2}{5 M}\right)\left(v.K_2 v.Eq_1+v.K_1 v.Eq_2\right)\right.\nonumber\\
&-&\left.\frac{22 M}{3m_2} v_1.K_1 v_1.Eq_1-\frac{22}{3} \left(v_2.K_1 v_1.Eq_2+v_1.K_2 v_2.Eq_1\right)\right]\nonumber\\
&+&\frac{G^2 m_1^2 m_2^2}{M}\left[-\frac{12}{5} r v^r \left(\dot{K}_1.Eq_1+\dot{Eq}_1.K_1\right)
+4 \left(r.\dot{K}_1 (v.Eq_1-v.Eq_2)+r.\dot{Eq}_1 (v.K_1-v.K_2)\right)\right.\nonumber\\
&-&\left.\frac{12}{5}\frac{M}{m_2} \left(r.\dot{K}_1 v.Eq_2+r.\dot{Eq}_1 v.K_2\right)
-\frac{34}{15} \frac{M}{m_2}\left(r.\dot{K}_2 v.Eq_1+r.\dot{Eq}_2 v.K_1\right)-\frac{4}{15} r.\dot{K}_1 r.\dot{Eq}_1\right]\nonumber\\
&+&G^2 m_1^2 m_2\left[\frac{12 m_2}{5 M}v^r -\frac{12}{5}v_1^r-\frac{74}{15}v_2^r\right]r 
\left(\dot{Eq}_1.K_2+\dot{K}_1.Eq_2\right)
-\frac45 \frac{G^2 m_1^2 m_2^2}{M} r^2 \dot{K}_1.\dot{Eq}_1\nonumber\\
&+&G^2 m_1^2 m_2 \left[r^2\left(\frac{2m_2}{5M}+\frac{19}{15}\right)\left(\dot{K}_2.\dot{Eq}_1+\dot{K}_1.\dot{Eq}_2\right)
+\frac{2}{15}\left(\frac{m_2}{M}-1\right)\left(r.\dot{K}_2 r.\dot{Eq}_1+r.\dot{Eq}_2 r.\dot{K}_1\right)\right]\nonumber\\
&+&\frac{G^3 m_1^2 m_2}{r^3}\left[m_2\!\left(\frac{224}{15}v^2-28{v^r}^2\right)\!-
\! m_1\left(\frac{8}{5}v^2+\frac{11}{6}r a_2^r\right)\!+\!\frac{Gm_1}{r}\!\left(\frac{268}{15}m_1+\frac{8}{3}m_2\right)\!\right]\!r.Eq_1\nonumber\\
&+&\frac{76}{5}\frac{G^4 m_1^3 m_2^2}{r^4}r.Eq_2+\frac{11}{3}G^2 m_1^3 v_1.a_1 v_1.Eq_1+\frac{52}{3}\frac{G^3 m_1^2 m_2^2}{r^2} v^r v.Eq_1\nonumber\\
&+&G^2 m_1^3\left[\frac{11}{3}+\frac{11}{2} v_1^2+\frac{55}{3}\frac{G m_2}{r}\right]a_1.Eq_1
-\frac{G^3 m_1 m_2}{r}\left(\frac{77}{6}m_1^2-4 m_2^2\right) a_2.Eq_1+{\cal O}(\varepsilon^2)\,.
\ee
Notice that only the last three lines contain ${\cal O}(\varepsilon)$ terms
(and thus a non-vanishing finite contribution to the regularized Lagrangian), whereas in the previous lines
$\vec{Eq}_{1,2}$ and $\vec{K}_{1,2}$ are combined to make the ${\cal O}(\varepsilon)$ contribution vanish.

\section{Results of double zero tricks}
\label{app:dztricks}
In deriving the previous results, the following double zero tricks have been used:
\subsection{From 2PN}
\be
&&{\cal L}^{(2\to 4PN)}_{dz}\rightarrow\frac{7 G^2 m_1^2 m_2}{4r^2}\left(2v_1^r v^r-v.v_1\right)
+\frac{7 G^3 m_1^2 m_2^2}{8 r^3}\nonumber\\
 &&+\frac{G^2 m_1^2 m_2}{r^2}\left\{a_1^r\left[\frac{1}{2}{v_2^r}^2+\frac{101}{16}v_1^2-\frac{101}{8}v_1.v_2 +\frac{7}{4}v_2^2\right]r-\frac{15}{2} r v_2.a_1 v_2^r-\frac{45}{8}r  v_1.a_2 v_1^r\right.\nonumber\\
&&+a_2^r\left[\frac{7}{16}{v_1^r}^2- \frac{15}{4} v_1^2\right] r
+\frac{45}{16}v_1^4+\frac{45}{8}(v_1.v_2)^2+\frac{15}{4}v_1^2 v_2^2-\frac{195}{16}v_1^2 v_1.v_2
-\frac{7}{16}v_2^2 {v_1^r}^2\nonumber\\
&&+\left.v_1^2\left(14v_1^r v_2^r-\frac{27}{8}{v_1^r}^2-\frac{15}{2}{v_2^r}^2\right)+
v_1.v_2\left(\frac{151}{16}v_1^r-\frac{97}{8}{v_2^r}\right)v_1^r+
\left(\frac{5}{4}v_1^r v_2^r-3{v_1^r}^2+\frac{7}{4}{v_2^r}^2\right){v_1^r}^2\right\}\nonumber\\
&&+\frac{7 G^3 m_1^3 m_2}{r^3}\left(v.v_1-3v_1^r v^r\right)
+\frac{G^3 m_1^2 m_2^2}{r^3}\left[14 v_1^2-\frac{63}{4} v_1.v_2 -\left(\frac{329}{8} v_1^r-\frac{77}{2} v_2^r\right) v_1^r\right]-\frac{63 G^4 m_1^3 m_2^2}{4r^4}
\nonumber\\
&&+\frac{G^2 m_1^2 m_2}{r^2}\left\{ v_2. a_1\left[ v_1^2\left(\frac{75}{8}v_1^r-\frac{15}{2} v_2^r\right)-\frac{15}{2}v_1. v_2v^r-\frac{15}{8}v_2^2 v_1^r+\left(\frac{45}{4}v_2^r-\frac{135}{16}v_1^r\right){v_1^r}^2\right]r \right.\nonumber\\
&&+v_1. a_1 \left(\frac{15}{2}v_1. v_2 v^r+\frac{15}{8}v_2^2 v_1^r\right) r+a_1^r\left[\frac{187}{32} v_1^4+ \frac{7}{2} \left(v_1. v_2\right)^2-7 v_1^2 v_1. v_2
+v_2^2\left(\frac{1}{8}v_2^r-\frac{11}{4}v_1^r\right)v_1^r\right.\nonumber\\
&&+v_1^2\left(\frac{1}{2} {v_2^r}^2+\frac{81}{8}v_1^r v_2^r-\frac{637}{32} {v_1^r}^2\right)+\left.v_1. v_2\left(10 {v_1^r}^2+v_1^rv_2^r-\frac{1}{2}{v_2^r}^2\right)+\left(\frac{21}{16}v_1^r-\frac{3}{4}v_2^r\right){v_1^r}^2 v_2^r\right]\nonumber\\
&&+a_2^r v_1^2\left(\frac{45}{8}{v_1^r}^2-\frac{15}{8}v_1^2\right)+\frac{75}{32}v_1^6-\frac{135}{32}v_1^4 v_1. v_2+\frac{15}{8}v_1^4 v_2^2
+v_1^4\left(\frac{315}{16}v_1^r v_2^r-\frac{555}{32}{v_1^r}^2-\frac{15}{4}{v_2^r}^2\right)\nonumber\\
&&+v_1^2 v_1. v_2\left(\frac{585}{32}v_1^r-\frac{45}{4}v_2^r\right)v_1^r-\frac{45}{8}v_1^2 v_2^2 {v_1^r}^2+v_1^2\left(\frac{453}{32}{v_1^r}^2-\frac{315}{8}v_1^r v_2^r+\frac{45}{2}{v_2^r}^2\right){v_1^r}^2\nonumber\\
&&+\left.\frac{87}{32}v_1. v_2{v_1^r}^4+\frac{261}{80}v^r {v_1^r}^5\right\}\nonumber\\
&&+\frac{G^3 m_1^3 m_2}{r^3}\left\{\frac{15}{4}v_2.a_1 \left(v_1^r+v_2^r\right) r
+a_1^r\left[14 v_1.v_2-\frac{41}{8} v_1^2 +\left(7 v_1^r-\frac{15}{4} v_2^r\right)v_2^r\right]r
\right.\nonumber\\
&&+\left.\frac{15}{8} \left(r a_2^r + v_1^2-v_2^2\right) v_1^2+v_1^2\left(\frac{45}{8} {v_2^r}^2-\frac{131}{8} {v_1^r}^2\right)
+\frac{43}{4} v_1.v_2{v_1^r}^2+\frac{215}{12} v^r {v_1^r}^3\right\}\nonumber\\
&&+\frac{G^3 m_1^2 m_2^2}{r^3}\left\{\frac{15}{4}v_2.a_1 v_1^r r
+a_1^r\left[\frac{1085}{32} v_1.v_2-\frac{965}{64} v_1^2 -\frac{525}{64} v_2^2-\frac{1}{8} {v_2^r}^2\right]r
-\frac{377}{64} v_1^4\right.\nonumber\\
&&-\frac{385}{32} \left(v_1.v_2\right)^2+\frac{467}{16} v_1^2 v_1.v_2-\frac{665}{64}  v_1^2 v_2^2
+v_1^2\left(\frac{2513}{64} {v_1^r}^2-\frac{4833}{32} v_1^r v_2^r+\frac{4617}{64} {v_2^r}^2\right)\nonumber\\
&&+\left.v_1.v_2\left(\frac{4621}{32}v_2^r-\frac{3269}{32}v_1^r\right)v_1^r
+\left(\frac{1415}{8} v_1^r v_2^r-\frac{15}{2} {v_1^r}^2-\frac{1313}{8} {v_2^r}^2\right){v_1^r}^2\right\}\nonumber\\
&&+\frac{G^4 m_1^3 m_2^2}{r^4}\left(\frac{2667}{16}v_1.v_2-\frac{1365}{16}v_1^2-\frac{399}{8}v_2^2+\frac{4949}{16}{v_1^r}^2-\frac{3423}{8}v_1^r v_2^r+\frac{2807}{16}{v_2^r}^2\right)\nonumber\\
&&+\frac{63 G^4 m_1^4 m_2}{4 r^4}\left(4v^r v_1^r -v.v_1\right)+\frac{385 G^5 m_1^3 m_2^3}{4 r^5}+\frac{1211 G^5 m_1^4 m_2^2}{16 r^5}\,.
\ee
\subsection{From 3PN}
\be
&&{\cal L}^{(3\to 4PN)}_{fin\,dz}\rightarrow\frac{G^2 m_1^2 m_2}{r^2}\left[\frac{27}{8}r v_2.a_1 v_2^r-2r v_1.a_2 v_2^r+\frac{27}{16}a_2^r\left(v_1^2+{v_1^r}^2\right)r
+a_1^r\left(\frac{1}{4}v_1^2-\frac{1}{2} v_1.v_2+v_2^2\right)r
\right.\nonumber\\
&&+\frac{1}{8} v_1^4+2(v_1.v_2)^2-\frac{7}{16}v_1^2v_1.v_2-\frac{27}{16}v_1^2 v_2^2
+v_1^2\left(\frac{1}{4}v_1^r v_2^r-\frac{13}{2} {v_1^r}^2+\frac{27}{8} {v_2^r}^2\right)
\nonumber\\
&&+v_1.v_2\left(\frac{191}{16}v_1^r-\frac{59}{8}v_2^r\right)v_1^r-\left.\frac{27}{16}v_2^2 {v_1^r}^2
+\left(\frac{25}{3}{v_1^r}^2- \frac{181}{12}v_1^r v_2^r+\frac{27}{4}{v_2^r}^2\right){v_1^r}^2\right]\nonumber\\
&&+\frac{G^3 m_1^3 m_2}{r^3}\left(\frac{29}{4}v.v_1-\frac{437}{12}v^r v_1^r\right)+\frac{G^3 m_1^2 m_2^2}{r^3}\left[\frac{1007}{144}v_1.v_2-\frac{409}{72}v_1^2+\left(\frac{281}{12}v_1^r-\frac{1013}{48}v_2^r\right) v_1^r\right]\nonumber\\
&&-\frac{7}{12}\frac{G^4 m_1^3 m_2^2}{r^4}+\frac{G_N^2 m_1^2 m_2}{r^2}\left\{v_1.a_1\left[v_1.v_2\left(\frac{245}{24}v_1^r-\frac{271}{24}v_2^r\right)
+v_2^2\left(\frac{407}{24}v_2^r-\frac{71}{8}v_1^r\right)\right]r\right.\nonumber\\
&+&v_2.a_1\left[v_1^2\left(\frac{355}{48}v_1^r -\frac{35}{24}v_2^r\right)+v_1.v_2\left(\frac{91}{24}v_2^r -\frac{139}{12}v_1^r\right)
+v_2^2\left(\frac{85}{8}v_1^r -\frac{407}{24}v_2^r\right)+\frac{61}{24}{v_1^r}^3\right.\nonumber\\
&-&\left.\frac{183}{16}{v_1^r}^2v_2^r-\frac{181}{48}v_1^r{v_2^r}^2+\frac{23}{12}{v_2^r}^3\right]r
+a_2^r v_1^2\left[\frac{23}{8}{v_2^r}^2-\frac{205}{48}v_1^rv_2^r-\frac{169}{32}{v_1^r}^2-\frac{215}{96}v_1^2\right]r\nonumber\\
&+&a_1^r\left[\frac{689}{192}v_1^4+\frac{17}{6}v_2^4-\frac{21}{8}v_1^2v_1.v_2+\frac{16}{3}v_1^2 v_2^2-\frac{75}{8}v_1.v_2 v_2^2
+v_1^2\left(\frac{967}{48}{v_1^r}^2-\frac{55}{6}v_1^r v_2^r-\frac{337}{96} {v_2^r}^2\right)\right.\nonumber\\
&+&\left.v_1.v_2 \left(\frac{67}{24}{v_2^r}^2-\frac{31}{6}v_1^r v_2^r-\frac{253}{12}{v_1^r}^2\right)+\left(\frac{439}{48}{v_1^r}^3+\frac{193}{48}{v_1^r}^2v_2^r-\frac{33}{16}v_1^r{v_2^r}^2-\frac{1}{4}{v_2^r}^3\right)v_2^r
\right.\nonumber\\
&+&\left.v_2^2\left(\frac{35}{6}v_1^r v_2^r-\frac{3}{2} {v_1^r}^2-\frac{2}{3} {v_2^r}^2\right)\right]r+v_1^4\left(\frac{233}{96}{v_1^r}^2-\frac{127}{96}v_1^r v_2^r
-\frac{635}{96}{v_2^r}^2\right)+\frac{457}{192}v_1^6+\frac{215}{96}v_1^4v_2^2\nonumber\\
&-&\frac{887}{192} v_1^4 v_1.v_2+v_1^2 v_1.v_2\left(\frac{481}{96}{v_2^r}^2+\frac{151}{24}v_1^rv_2^r-\frac{399}{32}{v_1^r}^2\right)
+v_1^2 v_2^2\left(\frac{169}{32}{v_1^r}^2+\frac{205}{48}v_1^rv_2^r-\frac{23}{8}{v_2^r}^2\right)\nonumber\\
&+&\left.v_1^2\left(\frac{67}{12}{v_1^r}^4+\frac{737}{24}{v_1^r}^3v_2^r-\frac{151}{12}{v_1^r}^2 {v_2^r}^2-\frac{99}{8}v_1^r{v_2^r}^3+\frac{23}{6}{v_2^r}^4\right)-\frac{91}{6}v_1.v_2 {v_1^r}^4-\frac{91}{5}v^r {v_1^r}^5\right\}\nonumber\\
&&+\frac{G^3 m_1^3 m_2}{r^3}\left\{a_1^r\left[\frac{161}{6}v_1.v_2-\frac{461}{18}v_1^2
-\frac{295}{24}v_2^2+\left(\frac{169}{12}v_1^r+ 5 v_2^r\right)v_2^r\right] r+\frac{215}{24} r a_2^r v_1^2
-\frac{101}{12} v_1^4\right.\nonumber\\
&&+v_2.a_1\left(\frac{215}{12}v_2^r-\frac{143}{6}v_1^r\right)r 
+\frac{139}{8}v_1^2 v_1.v_2-\frac{215}{24}v_1^2 v_2^2
+v_1^2\left(\frac{973}{16}{v_1^r}^2-\frac{2483}{24}v_1^r v_2^r+\frac{1349}{24}{v_2^r}^2\right)
\nonumber\\
&&\left.-\frac{217}{16}v_1.v_2 {v_1^r}^2-\frac{733}{48}v^r {v_1^r}^3\right\}
+\frac{G^3 m_1^2 m_2^2}{r^3}\left[a_1^r\left(\frac{4705}{288}v_1^2
-\frac{571}{9}v_1.v_2+\frac{89}{3}v_2^2-\frac{547}{96}{v_2^r}^2\right)r \right.\nonumber\\
&&-\frac{671}{144}r v_2.a_1v_1^r+\frac{6101}{288}v_1^4+\frac{1087}{36}\left(v_1.v_2\right)^2-\frac{21437}{288}v_1^2 v_1.v_2
+\frac{1471}{72}v_1^2 v_2^2+v_1.v_2\left(\frac{3005}{96} v_1^r-\frac{4097}{48} v_2^r\right)v_1^r\nonumber\\
&&\left.+v_1^2\left(\frac{4137}{32}v_1^r v_2^r-\frac{4025}{96}{v_1^r}^2-\frac{4019}{96}{v_2^r}^2\right)
+\left(\frac{30349}{288}v_1^r v_2^r-\frac{9023}{144}{v_1^r}^2-\frac{1589}{32}{v_2^r}^2\right) {v_1^r}^2\right]\nonumber\\
&&+\frac{G^4 m_1^3 m_2^2}{r^4}\left(\frac{487}{72}v_2^2-\frac{2009}{72}v_1.v_2-v_1^2
+\frac{1475}{144}{v_1^r}^2+\frac{1067}{18}v_1^r v_2^r-\frac{643}{6}{v_2^r}^2\right)\nonumber\\
&&+\frac{G^4 m_1^4 m_2}{r^4}\left(\frac{536}{3}v^r v_1^r-30v.v_1\right)
-\frac{13}{8}\frac{G^5 m_1^3 m_2^3}{r^5}+\frac{365}{36}\frac{G^5 m_1^4 m_2^2}{r^5}
\nonumber\\
&&\ds -2(d-3){\cal L}_{pole,dz}^{(3\to 4PN)}\log{r}\,,
\ee
with
\allowdisplaybreaks
\be
&&(d-3){\cal L}^{(3\to 4PN)}_{pole,dz}=
\frac{22 G^3 m_1^3 m_2}{3r^3}\left(v.v_1-3v_1^r v^r\right)
+\frac{11 G^3 m_1^3 m_2}{r^3}\left(v.v_1-\frac{11}{3}v_1^r v^r\right)
\nonumber\\
&&-\frac{11 G^4 m_1^3 m_2^2}{3r^4}-\frac{11 G^4 m_1^3 m_2^2}{r^4}\nonumber\\
&&+\left\{\frac{11G^3 m_1^3 m_2}{3r^3}\left[v_2.a_1 \left(8v_2^r-6v_1^r\right) r
+a_1^r\left(8v_1.v_2-7 v_1^2-2 v_2^2 \right)r+4r a_2^r v_1^2-3v_1^4\right.\right.\nonumber\\
&&+\left.7v_1^2 v_1.v_2-4v_1^2 v_2^2+v_1^2\left(6 {v_1^r}^2-21v_1^r v_2^r+12 {v_2^r}^2\right)
+3v_1.v_2{v_1^r}^2+5 v^r {v_1^r}^3\right]\nonumber\\
&&+\frac{11G^3 m_1^2 m_2^2}{3r^3}\left[4v_2.a_1 v_1^r r
+a_1^r\left(2v_1.v_2-v_1^2+v_2^2 \right)r-3 r a_1^r {v_2^r}^2 +v_1^4-2(v_1.v_2)^2\right.\nonumber\\
&&+\left.\left.2v_1^2 v_1.v_2-v_1^2 v_2^2+v_1^2\left(3 {v_2^r}^2-6 {v_1^r}^2\right)
+v_1.v_2 v_1^r\left(6v_2^r -3v_1^r\right)+5 v^r {v_1^r}^3\right]\right\}\nonumber\\
&&+\left\{\frac{11G^4 m_1^3 m_2^2}{3r^4}
\left(18 v_1.v_2-10 v_1^2-6 v_2^2+40 {v_1^r}^2-54 v_1^r v_2^r+17{v_2^r}^2\right)\right.\nonumber\\
&&+\left.\frac{88G^4 m_1^4 m_2}{3r^4}\left(4v^r v_1^r-v.v_1\right)\right\}
+\left\{\frac{88G^5 m_1^3 m_2^3}{3r^5}+\frac{110G^5 m_1^4 m_2^2}{3r^5}\right\}\,.
\ee

\subsection{From 4PN}
\be
&&{\cal L}^{4PN}_{G,dz}=\frac{G^2 m_1^2 m_2}{r^2}\left\{v_1. a_1\left[v_1. v_2\left(\frac{65}{48}v_1^r+\frac{137}{48}v_2^r\right)
-v_2^2 \left(\frac{81}{32}v_1^r+\frac{665}{96}v_2^r\right)\right]r\right.\nonumber\\
&&+v_2. a_1\left[v_1^2\left(\frac{95}{96}v_2^r-\frac{65}{96} v_1^r\right)+ v_1. v_2\left(\frac{103}{48}v_1^r+\frac{25}{48}v_2^r\right)
+v_2^2 \left(\frac{113}{32}v_1^r+\frac{665}{96}v_2^r\right)\right.\nonumber\\
&&+\left.\frac{557}{96}{v_1^r}^3+\frac{39}{32}{v_1^r}^2v_2^r-\frac{121}{96}v_1^r{v_2^r}^2-\frac{37}{32}{v_2^r}^3\right]r\nonumber\\
&&+a_1^r\left[\frac{185}{384}v_1^4+\frac{23}{16}\left(v_1. v_2\right)^2+\frac{89}{192}v_2^4-\frac{13}{16}v_1^2 v_1. v_2
-\frac{11}{96} v_1^2v_2^2+\frac{27}{16}v_1. v_2 v_2^2\right.\nonumber\\
&&+v_1^2\left(\frac{499}{96} v_1^r v_2^r-\frac{673}{192}{v_1^r}^2
-\frac{49}{192} {v_2^r}^2\right)+v_1. v_2\left(\frac{52}{3}{v_1^r}^2-\frac{73}{12}v_1^r v_2^r-\frac{19}{24} {v_2^r}^2\right)\nonumber\\
&&+\left.v_2^2\left(\frac{9}{8}{v_1^r}^2-\frac{71}{24}v_1^r v_2^r+\frac{5}{12}{v_2^r}^2\right)
+v_2^r\left(\frac{805}{96}{v_1^r}^2v_2^r+\frac{1}{6}v_1^r{v_2^r}^2-\frac{375}{16}{v_1^r}^3-\frac{7}{192} {v_2^r}^3\right)\right]r\nonumber\\
&&+a_2^r v_1^2\left[\frac{419}{384}v_1^2+\frac{147}{64}{v_1^r}^2-\frac{121}{96} v_1^r v_2^r-\frac{111}{64} {v_2^r}^2\right]r
-\frac{53}{384}v_1^6+\frac{59}{48}v_1^4 v_1. v_2-\frac{419}{384}v_1^4v_2^2\nonumber\\
&&+v_1^4\left(\frac{149}{96}{v_2^r}^2+\frac{205}{96}v_1^r v_2^r-\frac{19}{48}{v_1^r}^2\right)
+v_1^2v _1. v_2\left(\frac{95}{32}{v_1^r}^2-\frac{281}{48}v_1^r v_2^r-\frac{53}{48}{v_2^r}^2\right)\nonumber\\
&&+v_1^2v_2^2\left(\frac{111}{64}{v_2^r}^2+\frac{121}{96}v_1^r v_2^r-\frac{147}{64}{v_1^r}^2\right)+\frac{2615}{192} v_1. v_2 {v_1^r}^4\nonumber\\
&&+\left.v_1^2\left(\frac{281}{24}{v_1^r}^2{v_2^r}^2-\frac{2443}{192}{v_1^r}^4-\frac{121}{12}{v_1^r}^3v_2^r-\frac{5}{24}v_1^r{v_2^r}^3-\frac{37}{16}{v_2^r}^4\right)+\frac{523}{32} v^r{v_1^r}^5\right\}+\nonumber\\
&&+\frac{G^3 m_1^3 m_2}{r^3}\left\{-a_1^r\left[\frac{61}{60} v_1^2 +\frac{571}{720} v_1.v_2+\frac{1549}{1440} v_2^2+\left(\frac{35}{8}v_1^r-\frac{49}{48} v_2^r\right)v_2^r\right]r
\right.\nonumber\\
&&-\frac{185}{96}r a_2^r v_1^2 r -v_2.a_1 \left(\frac{487}{144}v_1^r+\frac{185}{48}v_2^r\right) r-\frac{401}{288} v_1^4-\frac{77}{144}  v_1^2 v_1.v_2+\frac{185}{96}  v_1^2 v_2^2\nonumber\\
&&+\left.v_1^2\left(\frac{227}{32} {v_1^r}^2+\frac{77}{48} v_1^r v_2^r -\frac{185}{32} {v_2^r}^2\right)-\frac{35}{12}v_1.v_2 {v_1^r}^2-\frac{175}{36}v^r {v_1^r}^3\right\}\nonumber\\
&&+\frac{G^3 m_1^2 m_2^2}{r^3}\left\{\frac{25}{24}v_2.a_1 v_1^r r
+a_1^r\left[\frac{1763}{2880} v_1^2+\frac{179}{480} v_1.v_2 -\frac{73}{320} v_2^2+\frac{3}{4} {v_2^r}^2\right]r
+\frac{161}{96} v_1^4\right.\nonumber\\
&&+\frac{283}{288} \left(v_1.v_2\right)^2-\frac{53}{96} v_1^2 v_1.v_2-\frac{269}{576}  v_1^2 v_2^2
+v_1^2\left(\frac{307}{960} {v_2^r}^2-\frac{3163}{960} {v_1^r}^2-\frac{173}{120} v_1^r v_2^r\right)\nonumber\\
&&+\left.v_1.v_2\left(\frac{1597}{120}v_1^r-\frac{1493}{480}v_2^r\right)v_1^r
+\left(\frac{6031}{576} {v_1^r}^2-\frac{5633}{288} v_1^r v_2^r+\frac{79}{8} {v_2^r}^2\right){v_1^r}^2\right\}\nonumber\\
&&+\frac{G^4 m_1^3 m_2^2}{r^4}\left(\frac{1339}{1440}v_1^2-\frac{19}{60}v_1.v_2-\frac{2033}{1440}v_2^2-\frac{2023}{1440}{v_1^r}^2-\frac{209}{80}v_1^r v_2^r -\frac{13}{32}{v_2^r}^2\right)\nonumber\\
&&+\frac{679 G^4 m_1^4 m_2}{ 720 r^3} \left(4 v^r v_1^r-v.v_1\right)+\frac{29 G^5 m_1^4 m_2^2}{ 30 r^5}-\frac{67 G_N^5 m_1^3 m_2^3}{160 r^5}\,,
\ee

\be
&&{\cal L}^{(4PN)}_{G^2fin,dz}=\frac{G^3_N m_1^3 m_2}{r^3}\left
[a_1^r\left(\frac{26}{75}v_2^2-\frac{25847}{3600}v_1^2-\frac{12599}{900}v_1.v_2+\frac{1207}{20}v_1^r v_2^r
-\frac{2767}{240}{v_2^r}^2\right) r\right.\nonumber\\
&&+v_2.a_1\left(\frac{8119}{360}v_2^r-\frac{3403}{120}v_1^r\right) r+\frac{8119}{720}r a_2^r v_1^2
-\frac{3907}{360}v_1^4+\frac{5311}{240}v_1^2 v_1.v_2-\frac{8119}{720}v_1^2 v_2^2
\nonumber\\
&&+\left.v_1^2\left(\frac{17671}{240}{v_1^r}^2-\frac{14813}{240}v_1^r v_2^r+\frac{6679}{240}{v_2^r}^2\right)
-\frac{3179}{80}v_1.v_2 {v_1^r}^2-\frac{3979}{48}v^r {v_1^r}^3\right]\nonumber\\
&&+\frac{G^3_N m_1^2 m_2^2}{r^3}\left[a_1^r\left(\frac{491}{50}v_1.v_2+\frac{1143}{200}v_1^2
-\frac{20813}{1800}v_2^2+\frac{103}{48}{v_2^r}^2\right)r-\frac{155}{36}r v_2.a_1v_1^r
-\frac{1711}{120}v_1^4\right.\nonumber\\
&&-\frac{6511}{360}\left(v_1.v_2\right)^2+\frac{3311}{80}v_1^2 v_1.v_2-\frac{6511}{720}v_1^2 v_2^2
+v_1^2\left(\frac{55783}{1200}{v_1^r}^2-\frac{24747}{400}v_1^r v_2^r+\frac{6311}{400}{v_2^r}^2\right)\nonumber\\
&&+\left.v_1.v_2\left(\frac{6311}{200} v_2^r-\frac{38341}{1200} v_1^r\right)v_1^r
+\left(\frac{2657}{240}v_1^r v_2^r-\frac{701}{80}{v_1^r}^2-\frac{277}{120}{v_2^r}^2\right) {v_1^r}^2\right]\nonumber\\
&&+\frac{G^4_N m_1^3 m_2^2}{r^4}\left[\frac{6277}{400}v_1^2-\frac{13021}{360}v_1.v_2+\frac{18073}{900}v_2^2
-\frac{13093}{240}{v_1^r}^2+\frac{24023}{300}v_1^r v_2^r-\frac{73331}{3600}{v_2^r}^2\right]\nonumber\\
&&+\frac{G^4 m_1^4 m_2}{r^3}\left(\frac{373}{1800}v.v_1+\frac{1907}{450}v^rv_1^r\right)
-\frac{18139}{1800}\frac{G^5_N m_1^3 m_2^3}{r^5}-\frac{6077}{900}\frac{G^5_N m_1^4 m_2^2}{r^5}\nonumber\\
&&-2(d-3){\cal L}^{(4PN)}_{G^2pole,dz}\log{r}\,,
\ee
with
\newpage
\be
&&(d-3){\cal L}^{(4PN)}_{G^2pole,dz}=-\left\{\frac{G^3 m_1^3 m_2}{r^3}
\left[a_1^r\left(\frac{19}{15}v_1^2 +\frac{92}{15} v_1.v_2+\frac{64}{15} v_2^2-28v_1^r v_2^r+3{v_2^r}^2\right)r
\right.\right.\nonumber\\
&&+3 r a_2^rv_1^2 +v_2.a_1 \left(\frac{26}{3} v_1^r+6v_2^r\right) r+\frac{2}{3}v_1^4+\frac{7}{3}v_1^2 v_1.v_2
-3 v_1^2 v_2^2+25 v_1.v_2 {v_1^r}^2+\frac{125}{3}v^r {v_1^r}^3\nonumber\\
&&\left.+v_1^2\left(9{v_2^r}^2-27  {v_1^r}^2-7 v_1^r v_2^r\right)\right\}
+\frac{G^3 m_1^2 m_2^2}{r^3}\left[a_1^r\left(-\frac{143}{15} v_1^2 +\frac{286}{15}v_1.v_2-\frac{11}{5} v_2^2-11 {v_2^r}^2\right)r\right.\nonumber\\
&&+\frac{44}{3}v_2.a_1  r v_1^r+\frac{17}{3} v_1^4-\frac{10}{3} \left(v_1.v_2\right)^2-\frac{2}{3} v_1^2 v_1.v_2
-\frac{5}{3} v_1^2 v_2^2+v_1^2\left(-\frac{246}{5}  {v_1^r}^2+\frac{272}{5} v_1^r v_2^r
-\frac{81}{5}{v_2^r}^2\right)\nonumber\\
&&\left.+v_1.v_2\left(\frac{217}{5}  v_1^r-\frac{162}{5}v_2^r\right)v_1^r+\left.\left(\frac{161}{3}  {v_1^r}^2-\frac{479}{3} v_1^r v_2^r+106{v_2^r}^2\right){v_1^r}^2\right]\right\}\nonumber\\
&&-\frac{1}{2}\left\{\frac{G^4 m_1^3 m_2^2}{r^4}\left(\frac{13}{15}v_1^2+\frac{76}{3}v_1.v_2-\frac{76}{5}v_2^2+34{v_1^r}^2-\frac{368}{5}v_1^r v_2^r+\frac{704}{15}{v_2^r}^2\right)\right.\nonumber\\
&&\left.-\frac{76}{15}\frac{G^4 m_1^4 m_2}{r^3}\left(4v^r v_1^r-v.v_1\right)\right\}
+\frac{4}{15}\frac{G^5 M m_1^3 m_2^2}{r^5}\,,
\ee
and finally
\be
&&{\cal L}^{(4PN)}_{G^3fin,dz}=\frac{G^4_N m_1^3 m_2^2}{r^4}\left[\left(\frac{4939}{450}+\frac{71}{16}\pi^2 \right)v_1^2+\left(\frac{17}{8}\pi^2-\frac{23834}{225}\right) v_1.v_2+\left(\frac{14243}{150}-\frac{105}{16}\pi^2 \right)v_2^2\right.\nonumber\\
&-&\left.\left(\frac{71}{4}\pi^2+\frac{32828}{225} \right){v_1^r}^2+\left(\frac{129086}{225}-\frac{17}{2}\pi^2 \right)v_1^r v_2^r+\left(\frac{105}{4}\pi^2-\frac{32086}{75}\right){v_2^r}^2 \right]\nonumber\\
&&+\frac{G^4_N m_1^4 m_2}{r^4}\left[\frac{517}{18}v.v_1-\frac{1106}{9}v^r v_1^r\right]+\frac{G^5_N m_1^4 m_2^2}{r^5}\left(\frac{105}{32}\pi^2-\frac{27827}{450}\right)
-\frac{G^5_N m_1^3 m_2^3}{r^5}\left(\frac{71}{32}\pi^2+\frac{4939}{900}\right)\nonumber\\
&&-3(d-3){\cal L}^{(4PN)}_{G^3pole,dz}\log{\bar r}\,,
\ee
with
\be
&&(d-3){\cal L}^{(4PN)}_{G^3pole,dz}=
\left\{\frac{G^4_N m_1^3 m_2^2}{r^4}\left[34 v_1^2-50 v_1.v_2+16 v_2^2-136{v_1^r}^2+200v_1^r v_2^r-64{v_2^r}^2 \right]\right.\nonumber\\
&&+\left.\frac{8}{3}\frac{G^4_N m_1^4 m_2}{r^4}\left[v.v_1-4v^r v_1^r\right]\right\}-
\left\{\frac{28}{3}\frac{G^5_N m_1^4 m_2^2}{r^5}+17\frac{G^5_N m_1^3 m_2^3}{r^5}\right\}\,.
\ee

\bibliography{ref4PN}

\end{document}